\documentclass{revtex4}
\usepackage{graphicx}

\def\duzomniejsze{<\kern-.7mm<}
\def\duzowieksze{>\kern-.7mm>}

\def\textbf#1{{\bf #1}}
\def\beq{\begin{equation}}
\def\eeq{\end{equation}}
\def\be{\begin{equation}}
\def\ee{\end{equation}}
\def\ben{\begin{eqnarray}}
\def\een{\end{eqnarray}}
\def\beqa{\begin{eqnarray}}
\def\eeqa{\end{eqnarray}}
\def\eea{\end{array}}
\def\bea{\begin{array}}
\newcommand{\bei}{\begin{itemize}}
\newcommand{\eei}{\end{itemize}}
\newcommand{\bee}{\begin{enumerate}}
\newcommand{\eee}{\end{enumerate}}

\def\>{\rangle}
\def\<{\langle}

\begin{document}

\title{Maximally entangled mixed states of two atoms trapped inside an optical cavity}

\begin{abstract}
In some off-resonant cases, the reduced density matrix of two atoms
symmetrically coupled with an optical cavity can very approximately
approach to maximally entangled mixed states or maximal Bell
violation mixed states in their evolution. The influence of phase
decoherence on the generation of maximally entangled mixed state is
also discussed.

\end{abstract}

\author{Shang-Bin Li$^{1,2}$}\email{stephenli74@yahoo.com.cn}, \author{Jing-Bo Xu$^1$}
\affiliation{1. Zhejiang Institute of Modern Physics and Department
of Physics,\\ Zhejiang University, Hangzhou 310027, People's
Republic of China} \affiliation{2. Shanghai research center of
Amertron-global,
Zhangjiang High-Tech Park, \\
299 Lane, Bisheng Road, No. 3, Suite 202, Shanghai, 201204, P.R.
China}

\maketitle

Quantum entanglement plays a crucial role in quantum information
processes \cite{1}. In the past few years, much attention has been
paid to the preparation of maximally entangled mixed states
\cite{2,3,28}. The properties of maximally entangled mixed state
have been studied by many authors \cite{4,5,6}. Maximally entangled
mixed states are those states that, for a given mixedness, achieve
the greatest possible entanglement. For two-qubit systems and for
various combinations of entanglement and mixedness measures, the
forms of the corresponding maximally entangled mixed states are
different \cite{6}. Using correlated photons from parametric
down-conversion, maximally entangled mixed states in the linear
entropy-concurrence plane have been created and characterized
\cite{2}. Generation and characterization of two-photon polarization
maximally entangled mixed states in the linear entropy-concurrence
plane has been also carried out, which is based on the peculiar
spatial characteristics of a high brilliance source of entangled
pairs \cite{3}. The preparation of maximally entangled mixed states
of two atoms asymmetrically on-resonance coupled with an optical
cavity has also been proposed \cite{28}. Recently, Clark and Parkins
\cite{26} have proposed a scheme to controllably entangle the
internal states of two atoms trapped in a high-finesse optical
cavity by employing quantum-reservoir engineering. By applying the
on-resonance atom-cavity couplings which are time-dependent,
Olaya-Castro et al. have also presented an efficient scheme for
controlled generation of entangled states of two atoms inside an
optical cavity\cite{27}. However, truly resonant coupling is not
available in realistic physical system. It is desirable to
investigate how the off-resonance coupling affects the preparation
of maximally entangled mixed states of two atoms. Based on our
previous analytical results in Ref.\cite{29}, in which the
entanglement behaviors of two atoms inside an optical cavity in the
presence of phase decoherence have been derived, we can easily
analyze the feasibility for preparing maximally entangled mixed
states in such a system. For keep this paper self-contained, we
briefly outline the basic contents about two two-level atoms inside
an optical cavity. Here, we investigate two two-level atoms
symmetrically coupling to single mode optical cavity and show that
in some off-resonant cases, the maximally entangled mixed states in
the plane of concurrence versus linear entropy of two atoms can be
very approximately generated. It is shown that the long time
entanglement behavior of two atoms is sensitive to the ratio of the
detuning and the coupling strength. The influence of the initial
mixedness of the atoms and phase decoherence is also analyzed.

Considering the system that two atoms are trapped inside single mode
optical cavity initially prepared in the vacuum state. The
Hamiltonian for the system can be given by \cite{9,10} ($\hbar=1$),
\be
H=\frac{\omega_0}{2}\sum^{2}_i\sigma^{(i)}_z+\omega{a}^{\dagger}a+g\sum^{2}_i(a\sigma^{(i)}_++a^{\dagger}\sigma^{(i)}_-),
\ee where $\sigma^{(i)}_z$, $\sigma^{(i)}_{\pm}$ ($i=1,2$) are
atomic operators, $\omega_0$ is atomic transition frequency, $g$ is
the coupling constant of individual atom to cavity field and $a$
($a^{\dagger}$) is the annihilation (creation) operator of cavity
field with frequency $\omega$. The generation of entangled state in
the system (1) in laboratory has been implemented \cite{10}. Various
modifications and generalizations of the system (1) have been
studied for preparing entangled states or realizing various kinds of
quantum information processes \cite{11,12,13,14}. It is assumed that
the cavity field are prepared initially in vacuum state $|0\rangle$,
and the atom 1 is prepared in the mixed state
$\lambda|e\rangle\langle{e}|+(1-\lambda)|g\rangle\langle{g}|$ and
the atom 2 is in the ground state $|g\rangle$, i.e., \be
\rho(0)=|0\rangle\langle0|\otimes[\lambda|e\rangle\langle{e}|+(1-\lambda)|g\rangle\langle{g}|]\otimes|g\rangle\langle{g}|.
\ee The time evolution of $\rho(t)$ can be derived as follows, \beqa
\rho(t)&=&\frac{\lambda}{8}[1+\frac{\Delta^2}{\Omega^2}+(1-\frac{\Delta^2}{\Omega^2})\cos\Omega{t}]|0\rangle\langle0|\otimes|B_+\rangle\langle{B_+}|
\nonumber\\
&&+\frac{g^2\lambda}{\Omega^2}[1-\cos\Omega{t}]|1\rangle\langle1|\otimes|gg\rangle\langle{gg}|+\frac{\lambda}{4}|0\rangle\langle0|\otimes|B_-\rangle\langle{B_-}|
\nonumber\\
&&+\frac{\sqrt{2}g\lambda}{2\Omega}[\frac{\Delta}{\Omega}(1-\cos\Omega{t})+i\sin\Omega{t}]|0\rangle\langle1|\otimes|B_+\rangle\langle{gg}|
\nonumber\\
&&+\frac{\sqrt{2}g\lambda}{2\Omega}[\exp(\frac{i\Omega{t}-i\Delta{t}}{2})-\exp(\frac{-i\Omega{t}-i\Delta{t}}{2})]|0\rangle\langle1|\otimes|B_-\rangle\langle{gg}|
\nonumber\\
&&+\frac{\lambda}{4}\{(1-\frac{\Delta}{\Omega})e^{i(\Omega+\Delta)t/2}+(1+\frac{\Delta}{\Omega})e^{-i(\Omega-\Delta)t/2}\}|0\rangle\langle0|\otimes|B_+\rangle\langle{B_-}|
\nonumber\\
&&+\frac{1-\lambda}{2}|0\rangle\langle0|\otimes|gg\rangle\langle{gg}|+h.c.,
\eeqa where $\Delta=\omega_0-\omega$ is the detuning between the
atoms and cavity field, $\Omega=(\Delta^2+8g^2)^{1/2}$, and
$|B_{\pm}\rangle=\frac{\sqrt{2}}{2}(|eg\rangle\pm|ge\rangle)$ are
the Bell states. By tracing out the degree of freedom of the cavity
field, we obtain the reduced density matrix $\rho_s(t)$ describing
the subsystem containing only two atoms, \beqa
\rho_s(t)&=&\frac{\lambda}{8}[1+\frac{\Delta^2}{\Omega^2}+(1-\frac{\Delta^2}{\Omega^2})\cos\Omega{t}]|B_+\rangle\langle{B_+}|
\nonumber\\
&&+\frac{\lambda}{4}|B_-\rangle\langle{B_-}|+\{\frac{g^2\lambda}{\Omega^2}[1-\cos\Omega{t}]+\frac{1-\lambda}{2}\}|gg\rangle\langle{gg}|
\nonumber\\
&&+\frac{\lambda}{4}\{(1-\frac{\Delta}{\Omega})e^{i(\Omega+\Delta)t/2}+(1+\frac{\Delta}{\Omega})e^{-i(\Omega-\Delta)t/2}\}|B_+\rangle\langle{B_-}|+h.c..
\eeqa

Firstly, we analyze the feasibility of preparing maximally entangled
mixed states of two atoms in this cavity QED system. The concurrence
\cite{18} is adopted to quantify the bipartite entanglement between
two atoms, and the linear entropy defined by
$M=\frac{4}{3}(1-{\mathrm{Tr}}\rho^2_s)$ of the reduced density
matrix is used to quantify the mixedness. In the situation with a
rational value of $\frac{\Delta}{\Omega}$, the evolving density
matrix is periodic. In the case with an irrational value of
$\frac{\Delta}{\Omega}$, the evolving state is not periodic. The
explicit analytical expression of the concurrence $C_s(t)$
characterizing the entanglement in $\rho_s(t)$ can be obtained as
\beqa C_s(t)&=&\lambda(A^2+B^2)^{\frac{1}{2}}, \nonumber\\
A&=&\frac{\Delta^2}{4\Omega^2}-\frac{1}{4}+\frac{1}{4}(1-\frac{\Delta^2}{\Omega^2})\cos\Omega{t},
\nonumber\\
B&=&\frac{1}{2}(1-\frac{\Delta}{\Omega})\sin(\Omega+\Delta)t/2-\frac{1}{2}(1+\frac{\Delta}{\Omega})\sin(\Omega-\Delta)t/2.
\eeqa In the case with $\lambda=1$, two atoms can be in the pure
states at some specific times. The entanglement characterized by
concurrence of those pure states are given by
$C=|\sin\frac{\Delta{k\pi}}{\Omega}|$ which are achieved at the
discrete times denoted by ${t}=2k\pi/\Omega$ ($k=1,2,...$). If
$\frac{\Delta}{\Omega}$ is a rational number, the series
$|\sin\frac{\Delta{k\pi}}{\Omega}|$ ($k=1,2,...$) have finite and
discrete values. While for the case that $\frac{\Delta}{\Omega}$ is
an irrational number, the series $|\sin\frac{\Delta{k\pi}}{\Omega}|$
($k=1,2,...$) have infinite numbers of values, and this series can
very approximately approach to any values between 0 and 1 according
to Hurwitz's theorem in number theory. It means pure two-qubit
states with any desired degree of entanglement can be very
approximately generated for those cases with the irrational values
of $\frac{\Delta}{\Omega}$.

In the large detuning limit, i.e., $g/|\Delta|\ll1$, the population
of the single mode cavity field will be very small in the time
evolution, which leads very small entanglement between the atoms and
the cavity field. Therefore the mixedness of the subsystem
containing two atoms is very small. In the small detuning limit,
i.e. $|\Delta|/g\ll1$ but not zero, and simultaneously
$\frac{\Delta}{\Omega}$ is an irrational number, the trajectories of
the reduced density operator of two atoms in the concurrence versus
linear-entropy plane exhibit a kind of "quasi-ergodic" property,
roughly speaking, where "quasi-ergodic" means there is no distinct
interspaces in the pattern formed by the trajectory of the evolving
state in the concurrence versus linear-entropy plane.

In Fig.1, the concurrence versus mixedness of two atoms are depicted
for different values of detuning. In the resonant case, the
concurrence of two atoms increases (decreases) with the increase
(decrease) of mixedness of their reduced density matrix. In the
resonant situation, the evolving reduced density matrix $\rho_s(t)$
in Eq.(4) can not become any one of the maximally entangled mixed
state in the plane of linear entropy-concurrence. Interestingly, in
the off-resonant case, part of the frontier of the concurrence
versus linear entropy can be very approximately reached by the
evolving reduced density matrix of two atoms. However, the region in
the frontier which can be approximately approached by the evolving
reduced density matrix reduces with the increase of the detuning.
Approximately, two atoms can acquire desired pure state concurrence
between 1 and 0 for both the small detuning case and large detuning
case in the precondition that $\frac{\Delta}{\Omega}$ is an
irrational number.

From Fig.2, we can understand the influence of initial mixedness of
two atoms on the entanglement and mixedness of their evolving reduce
density matrix. It is shown that the range of the frontier which can
be approximately approached is reduced when the initial mixedness of
the atoms increases. For the case with $\lambda=0.6$, two atoms can
evolve into a state with smaller mixedness than their initial state
which is different from other cases with $\lambda=0.9$ and
$\lambda=0.7$. One can also find that the patterns formed by the
trajectories are mirror symmetric with the horizontal axis labeled
by the half of the concurrence of maximally entangled mixed state
corresponding to the initial linear entropy.

\begin{figure}
\centerline{\includegraphics[width=1.5in]{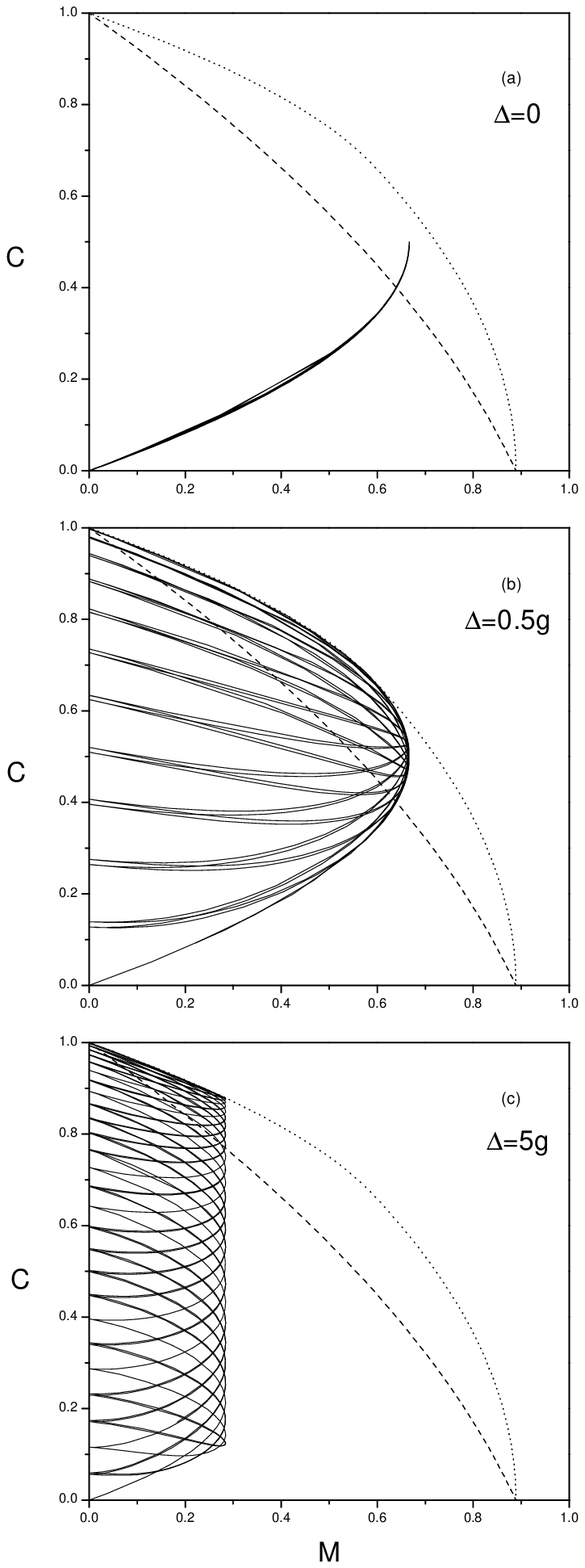}} \caption{The
concurrence versus mixedness of two atoms are depicted. The
trajectory is chosen from the scaled time $gt\in[0,50]$. The dash
line and dot line in (a), (b) and (c) represent the Werner state and
the maximally entangled mixed state (the frontier of the concurrence
versus linear entropy) respectively. (a) $\Delta=0$; (b)
$\Delta=0.5g$; (c) $\Delta=5g$. In three cases, $\lambda=1$.}
\end{figure}

\begin{figure}
\centerline{\includegraphics[width=1.5in]{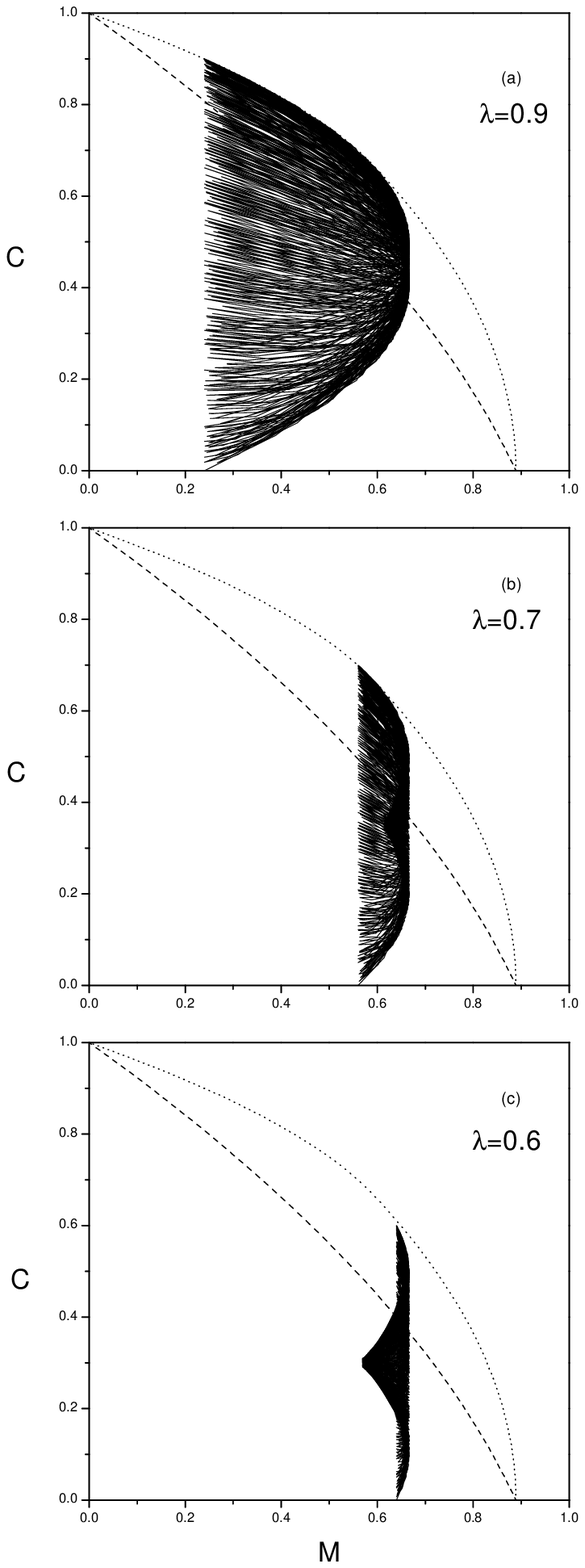}} \caption{The
concurrence versus mixedness of two atoms are displayed in the cases
in which one of the atoms is initially in three different mixed
states: (a) $\lambda=0.9$; (b) $\lambda=0.7$; (c) $\lambda=0.6$. The
trajectories are also chosen from the scaled time $gt\in[0,500]$.
The dash line and dot line in (a), (b) and (c) represent the Werner
state and the maximally entangled mixed state respectively. It is
shown that two atoms can approximately approach part of the
maximally entangled mixed states though the range of the frontier
which can be approached is reduced when the initial mixedness of the
atom increases. In three cases, $\Delta=0.5g$. }
\end{figure}
\begin{figure}
\centerline{\includegraphics[width=1.5in]{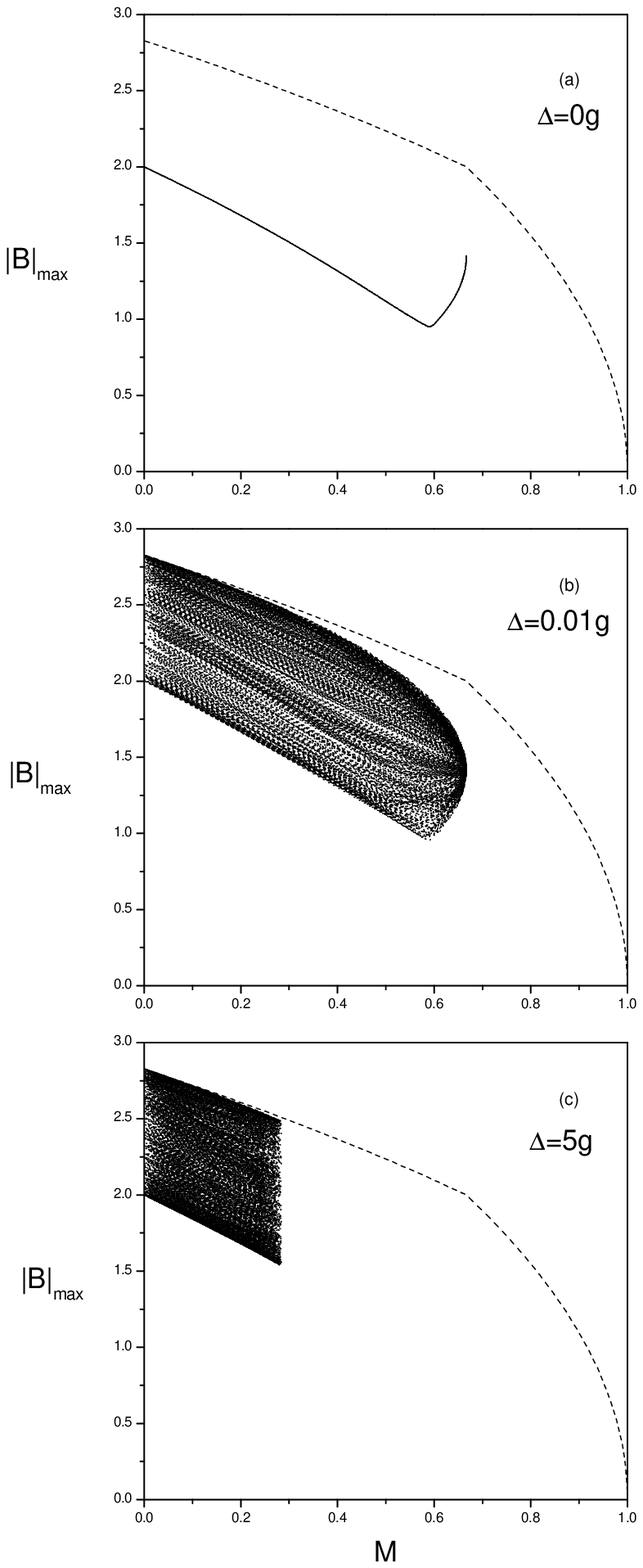}} \caption{The
maximal Bell violation $|B|_{max}$ versus the mixedness of two atoms
is displayed for three different values of the detuning: (a)
$\Delta=0$; (b) $\Delta=0.01g$; (c) $\Delta=5g$. The trajectory is
chosen from the scaled time $gt\in[0,500]$. The dash line represents
the frontier of maximal Bell violation versus the linear entropy,
namely, for a given linear entropy, the maximal value of $|B|_{max}$
of two atoms can not exceed the dash line. In three cases,
$\lambda=1$.}
\end{figure}

Bell's inequality test with entangled atoms inside a cavity have
been extensively studied \cite{19}. The most commonly discussed Bell
inequality is the CHSH inequality \cite{20,21}. The CHSH operator
reads \be
\hat{B}=\vec{a}\cdot\vec{\sigma}\otimes(\vec{b}+\vec{b^{\prime}})\cdot\vec{\sigma}
+\vec{a^{\prime}}\cdot\vec{\sigma}\otimes(\vec{b}-\vec{b^{\prime}})\cdot\vec{\sigma},
\ee where $\vec{a},\vec{a^{\prime}},\vec{b},\vec{b^{\prime}}$ are
unit vectors. In the above notation, the Bell inequality reads \be
|\langle\hat{B}\rangle|\leq2. \ee The maximal amount of Bell
violation of a state $\rho$ is given by \cite{22} \be
|{\mathcal{B}}|_{max}=2\sqrt{\kappa+\tilde{\kappa}},\ee where
$\kappa$ and $\tilde{\kappa}$ are the two largest eigenvalues of
$T^{\dagger}_{\rho}T_{\rho}$. The matrix $T_{\rho}$ is determined
completely by the correlation functions being a $3\times3$ matrix
whose elements are
$(T_{\rho})_{nm}={\mathrm{Tr}}(\rho\sigma_{n}\otimes\sigma_{m})$.
Here, $\sigma_1\equiv\sigma_x$, $\sigma_2\equiv\sigma_y$, and
$\sigma_3\equiv\sigma_z$ denote the usual Pauli matrices. The
quantity $|\mathcal{B}|_{max}$ is called as the maximal violation
measure, which indicates the Bell violation when
$|{\mathcal{B}}|_{max}>2$ and the maximal violation when
$|{\mathcal{B}}|_{max}=2\sqrt{2}$. For the density operator $\rho_s$
in Eq.(4), $\kappa+\tilde{\kappa}$ can be written as follows \be
\kappa+\tilde{\kappa}=\varsigma+\max[\varsigma,\zeta], \ee where
\beqa \varsigma&=&\frac{4g^4}{\Omega^4}(1-\cos\Omega{t})^2
\nonumber\\
&&+\frac{1}{4}[(1-\frac{\Delta}{\Omega})\sin\frac{(\Omega+\Delta)t}{2}-(1+\frac{\Delta}{\Omega})\sin\frac{(\Omega-\Delta)t}{2}]^2
\nonumber\\
\zeta&=&(\frac{\Delta^2+4g^2}{\Omega^2}+\frac{4g^2}{\Omega^2}\cos\Omega{t})^2
\eeqa in the case with $\lambda=1$. In Ref.\cite{6}, the analytical
form of the mixed states which possess the maximal value of
$|{\mathcal{B}}|_{max}$ of two qubits for a given linear entropy has
been derived. Here, part of the frontier of the maximal Bell
violation versus the linear entropy can also be very approximately
approached by the evolving state of two atoms (see Fig.3b). In
Fig.3a, our calculations show that two atoms can not violate the
Bell-CHSH inequality in the resonant case, though two atoms could
get entangled. While in the off-resonant case, the bell violation of
the atom 1 and the atom 2 can emerge in their long-time evolution,
even though the detuning $\Delta$ is very very small.

\begin{figure}
\centerline{\includegraphics[width=1.5in]{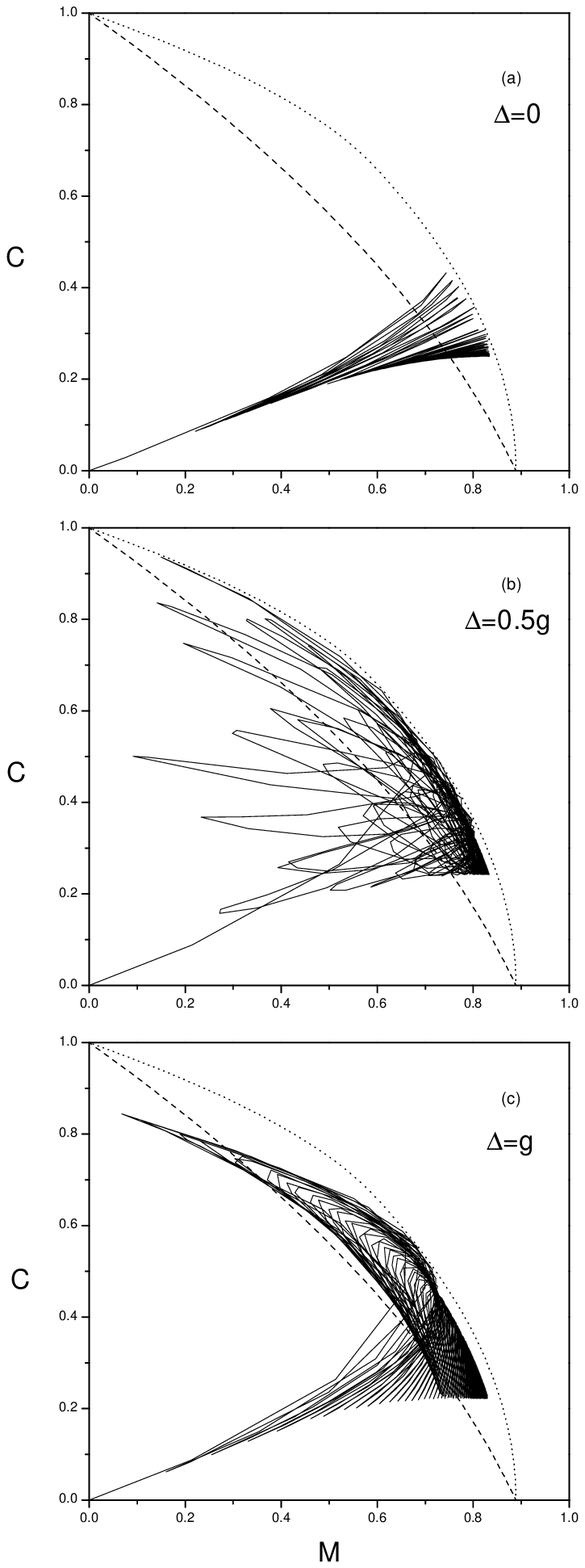}} \caption{We
display the concurrence versus mixedness of two atoms in the
presence of phase decoherence. The trajectories are chosen from the
scaled time $gt\in[0,500]$. The dash line and dot line in (a), (b)
and (c) represent the Werner state and the maximally entangled mixed
state, respectively. (a) $\Delta=0$; (b) $\Delta=0.5g$, it can be
observed that an appropriate detuning and decoherence rate can make
two atoms possess of ability to approach the wider region of the
frontier; (c) $\Delta=g$. In three cases, $\gamma=0.01/g$ and
$\lambda=1$.}
\end{figure}

If the pure phase decoherence mechanism is considered, the master
equation governing the time evolution of the system under the
Markovian approximation is given by \cite{15,16} \be
\frac{d\rho}{dt}=-i[H,\rho]-\frac{\gamma}{2}[H,[H,\rho]], \ee where
$\gamma$ is the phase decoherence rate. The explicit analytical
expression of the concurrence $C_{\gamma}(t)$ characterizing the
entanglement of two atoms in the presence of phase decoherence can
be obtained as
\beqa C_{\gamma}(t)&=&\lambda(A_{\gamma}^2+B_{\gamma}^2)^{\frac{1}{2}}, \nonumber\\
A_{\gamma}&=&\frac{\Delta^2}{4\Omega^2}-\frac{1}{4}+\frac{1}{4}(1-\frac{\Delta^2}{\Omega^2})\cos\Omega{t}\exp(-\frac{\gamma{t}}{2}\Omega^2),
\nonumber\\
B_{\gamma}&=&\frac{1}{2}(1-\frac{\Delta}{\Omega})\sin(\Omega+\Delta)t/2\exp[-\frac{\gamma{t}}{8}(\Omega+\Delta)^2]-\frac{1}{2}(1+\frac{\Delta}{\Omega})\sin(\Omega-\Delta)t/2\exp[-\frac{\gamma{t}}{8}(\Omega-\Delta)^2],
\eeqa if the system is initially in the same state as $\rho(0)$ in
Eq.(2). From Eq.(12), we can easily know that the phase decoherence
does not completely destroy the entanglement but generate a
stationary entangled state of two atoms. The concurrence
$C_{\gamma}(t)$ is not larger that 0.5 in the resonant case. The
entanglement of stationary state decreases with the increase of the
detuning. The phase decoherence changes trajectories in the plane of
concurrence versus linear entropy of the evolving state and makes
the trajectories become chaotic. In Fig.4, we display the
concurrence versus mixedness of two atoms in the presence of phase
decoherence. The evolving reduced density matrix of two atoms can
approximately approach to wider region of the maximally entangled
mixed states, if both the ratio $\Delta/g$ and the decoherence rate
$\gamma$ are appropriate.

In summary, we have investigated a possible scheme for generating
the maximally entangled mixed state of two atoms which are
symmetrically coupled to a single mode optical cavity field. It is
shown that two atoms can not achieve the maximally entangled mixed
state in the resonant case. In the off-resonant case, the reduced
density matrix of two atoms can approximately approach to the
maximally entangled state in their evolution. The distinct roles of
the rational values or irrational values of $\frac{\Delta}{\Omega}$
in the long-time behaviors of entanglement and mixedness of two
atoms have been clarified. The influence of the phase decoherence
and the initial mixedness of the atoms is also discussed. These
results presented here maybe have potential applications in the
domain of quantum information and quantum communication and in the
field dealing with the fundamental tests of quantum mechanics.

\end{document}